\documentclass[preprint,12pt]{article}

\usepackage{color}
\usepackage{xcolor}
\usepackage{ulem}
\usepackage{cancel}
\usepackage{amssymb}

\newtheorem{theorem}{Theorem}

\newtheorem{remark}[theorem]{Remark}

\begin{document}

\title{Some comments on the Butler-Volmer equation for modeling Lithium-ion batteries}

\author{A.M. Ramos\thanks{e-mail: angel@mat.ucm.es} \\
Instituto de Matem\'atica Interdisciplinar \\
\& Departamento de Matem\'atica Aplicada \\
Universidad Complutense de Madrid \\
%Plaza de Ciencias 3, Madrid, 28040, Spain \\
\and \\
C.P. Please \\
Mathematical Institute, University of Oxford} % \\
%Andrew Wiles Building, Radcliffe Observatory Quarter \\
%Woodstock Road, Oxford OX2 6GG, United Kingdom}

\maketitle

\begin{abstract}
  In this article the Butler-Volmer equation used in describing
  Lithium-ion (Li-ion) batteries is discussed. First, a complete
  mathematical model based on a macro-homogeneous approach developed
  by Neuman is presented.  Two common mistakes found in the literature
  regarding a sign in a boundary conditions and the use of the
  transfer coefficient are mentioned. The paper focuses on the form of
  the Butler-Volmer equation in the model. It is shown how practical
  problems can be avoided by taking care in the form used, particularly
  to avoid difficulties when the solid particle in the electrodes
  approach a fully charged or discharged state or the electrolyte gets
  depleted. This shows that the open circuit voltage and the exchange
  current density must depend on the lithium concentration in both the
  solid and the electrolyte in a particular way at the extremes of the
  concentration ranges.
\end{abstract}
%\begin{center}
Keywords: Lithium-ion batteries, Butler-Volmer equation, mathematical model.
%\end{center}

\section{Introduction}
Lithium-ion (Li-ion) batteries have become very popular in the last
years as efficient energy storage devices.  A mathematical model
demonstrating the role of the key factors in the battery operation can
be very helpful for the design and optimization of new models and also
for the real time control of its performance.

Based on a macro-homogeneous approach developed by Neuman
\cite{1973Ne}, several mathematical models have been developed for
these purposes
(cf. \cite{1993DoFuNe,1994FuDoNe,2002GoEtAl,2005FaPl,2006SmWa,2006-b-SmWa,2007SmRaWa,2010ChEtAl,2012KiSmIrPe},
which include the main physics present in charge/discharge
processes. A complete mathematical model is presented in Section
\ref{themodel} including a system of boundary value problems for the
conservation of Lithium and conservation of charge in the solid and
electrolyte phases. In that section we remark that several authors
have included a boundary condition that is not correct (see
\cite{2014Ra} and Remark \ref{rmswbc}). Although the authors have
probably used the correct boundary condition when solving the model
numerically the reader should be aware of the right choice. A second
common mistake involving the role of the transfer coefficient in the
equation for conservation of charge in the electrolyte is also pointed
out (see Remark \ref{semist}).

Once we have presented the model, Section \ref{sdbve} is devoted to
the Butler-Volmer equation.  It is shown that care must be taken in
the form of this equation especially in the limiting cases where the
electrodes approach a fully charged or discharged state or the local
electrolyte concentration reduces substantially. In such cases it is
necessary to ensure the equation allows particles to discharge or
charge but only in direction that ensures concentrations remain in the
physically relevant regime.

\section{Mathematical model} \label{themodel}

\subsection{Generalities}

Let us assume we have a general binary electrolyte (i.e. a single salt
composed of one kind of cation and one kind of anion) conducting
electricity and there is no convection involved in the process. In
order to give a general idea of the equations involved in the model
detailed in Section \ref{ssafmm} below, we use the infinite dilute
solution theory (we could also use the concentrated solution theory,
see \cite{1973Ne}, which is more complex and accurate but leads to
equations of the same type).

The accumulation of ions is due to the net flux and the production, i.e.
\begin{equation} \label{mbfei}
\frac{\partial c_i}{\partial t} = -\nabla \cdot N_{i} + R_i  \ \ \ \ \ (i\in \{ +,-\}),
\end{equation}
where $R_i$ (mole m$^{-3}$ s$^{-1}$) is the reaction rate and $N_i$ is
the flux (mole m$^{-2}$ s$^{-1}$) of anions ($N_-$) and cations
($N_+$) in the electrolyte. This flux is due to migration and
diffusion and is given by
\begin{equation} \label{foie}
N_{i} = -z_i u_i F c_i \nabla \phi_{\rm e} - D_i \nabla c_i \ \ \ \ \ (i\in \{ +,-\}),
\end{equation}
where $c_+, c_-$ are the molar concentration (mol m$^{-3}$) of cations
and anions, $z_+, z_-$ are the number of protons charges carried by a
cation and by an anion, $u_+, u_-$ are the mobility of cations and
anions (m$^2$ mole J$^{-1}$ s$^{-1}$) and $D_+, D_-$ are the diffusion
coefficient of cations and anions (m$^2$ s$^{-1}$).

We assume the solution is electrically neutral, i.e.
\begin{equation} \label{sien}
z_+c_+ + z_-c_-=0
\end{equation}

Let $\nu_+, \nu_-$ be the number of cations and anions produced by the dissociation of one molecule of electrolyte, i.e.
$$
z_+\nu_+ + z_-\nu_-=0.
$$
and
$$
c_{\rm e} =\frac{c_+}{\nu_+} = \frac{c_-}{\nu_-} \ \ \ \mbox{(so called, concentration of the electrolyte)}.
$$
Assuming
$u_+$ and $u_-$ are constant, if we multiply (\ref{mbfei}) by $\frac{1}{\nu_\pm}$ we get
$$
\frac{\partial c}{\partial t} - z_+ u_+ F \nabla \cdot \left( c \nabla \phi_{\rm e}\right)  - \nabla \cdot \left( D_+ \nabla c \right)  = \frac{R_+}{\nu_+},
$$
$$
\frac{\partial c}{\partial t} - z_- u_- F \nabla \cdot \left( c \nabla \phi_{\rm e}\right)  - \nabla \cdot \left( D_- \nabla c \right)  = \frac{R_-}{\nu_-}.
$$
Subtraction of these two equaitons gives
$$
(z_+ u_+  - z_- u_-) F \nabla \cdot \left( c \nabla \phi_{\rm e}\right)  + \nabla \cdot \left( (D_+ -D_-) \nabla c \right)  = \frac{R_-}{\nu_-} - \frac{R_+}{\nu_+},
$$
which can be used to eliminate $\nabla \phi_{\rm e}$ from either of the previous equations in order to get the equation for the conservation of ions
\begin{equation} \label{ecoi}
\frac{\partial c_{\rm {e}}}{\partial t} -  \nabla \cdot \left( D_{\rm e} \nabla c_{\rm {e}} \right)
=  R,
\end{equation}
where
$$
D_{\rm e}= \frac{z_+ u_+D_- - z_-u_- D_+}{z_+ u_+  - z_- u_-}, \ \mbox{ and } \ \ R= \frac{z_+ u_+\frac{R_-}{\nu_-} - z_- u_-\frac{R_+}{\nu_+}}{z_+ u_+  - z_- u_-}.
$$
The diffusion $D_{\rm e}$ and reaction $R$ coefficients represent a compromise between the diffusion and the reaction coefficients of the anion and the cation.

Regarding the electrical current in the electrolyte, we have that
$$
i_{\rm e}=F(z_+N_+ + z_-N_-),
$$
where $i_{\rm e}$ is current density (A m$^{-2}$). Then, from (\ref{mbfei})--(\ref{ecoi}) it can be deduced that
\begin{equation} \label{ecoc1}
-\frac{\nabla \cdot i_{\rm e}}{z_+\nu_+ F}  = \frac{R_-}{\nu_-} - \frac{R_+}{\nu_+}
\end{equation}
and, using the Nernst-Einstein equation, $D_i = RTu_i$, that
\begin{equation} \label{ecoc2}
i_{\rm e} = -\kappa \nabla \phi_{\rm e}  - \frac{RT\kappa }{F} \left( \frac{t_+}{z_+}  + \frac{t_-}{z_-} \right)  \nabla \ln c_{\rm  e},
\end{equation}
where $\kappa = F^2\left( (z_+)^2u_+c_+ + (z_-)^2u_-c_-\right)$ is  electrical conductivity of the electrolyte and
$$
t_+=  \frac{z_+u_+}{z_+u_+  - z_-u_-} , \ \ t_-= \frac{-z_-u_-}{z_+u_+  - z_-u_-}=1-t_+
$$
are the so-called transference numbers. Hence, from (\ref{ecoc1}) and (\ref{ecoc2}) we obtain the equation for the conservation of charge
\begin{equation} \label{ecoc}
-\nabla \cdot \left( \kappa \nabla \phi_{\rm e} \right)  - \frac{RT}{F} \left( \frac{t_+}{z_+}  + \frac{t_-}{z_-} \right)  \nabla \cdot \left( \kappa \nabla \ln c_{\rm  e} \right) = z_+\nu_+ F \left( \frac{R_+}{\nu_+} - \frac{R_-}{\nu_-} \right) .
\end{equation}

\begin{remark} \label{resce}
For Li$^+$ batteries with $z_+=1$, $z_- = -1$, $\nu_+=\nu_-=1$ and $R_-=0$ (i.e. the cation is the only ion reacting at the electrode) the following equations are satisfied in the electrolyte:
$$
\frac{\partial c_{\rm e}}{\partial t} -  \nabla \cdot \left( D_{\rm e} \nabla c_{\rm  e} \right)
= \frac{1-t_+}{F}j^{\rm Li} \ \mbox{ (conservation of ions)}
$$
$$
-\nabla \cdot \left( \kappa \nabla \phi_{\rm e} \right)  - \frac{RT}{F} \left( 2t_+-1 \right)  \nabla \cdot \left( \kappa \nabla \ln c_{\rm e} \right) = j^{\rm Li}
\ \mbox{ (conservation of charge)}.
$$
We have used that, due to the Faraday's laws of electrolysis,
$
R_+= \frac{1}{F} j^{\rm Li},
$
where $j^{\rm Li}=\frac{\partial i_{\rm e}}{\partial x}$ is the reaction current (A m$^{-3}$) resulting in production or consumption of Li$^+$.
\end{remark}

\subsection{The complete model} \label{ssafmm}

A typical Li-ion battery cell has three regions: A porous negative
electrode, a porous positive electrode and an electron-blocking
separator. In all these regions there is an electrolyte containing
various charge species, including lithium, that can move all through
the cell in response to an electrochemical potential gradient.

A 1D electrochemical model is considered for the evolution of the Li
concentration $c_{\rm e}(x,t)$ (mol m$^{-3}$) and the electric
potential $\phi_{\rm e}(x,t)$ (V) in the electrolyte and the electric
potential $\phi_{\rm s}(x,t)$ (V) in the solid electrodes, along the
$x$--direction, with $x\in (0,L)$ and $L=L_1+\delta +L_2$ being the
cell width (m). We assume that $(0,L_{1})$ corresponds to the negative
electrode, ($L_1,L_1+\delta)$ corresponds to the separator and
$(L_1+\delta,L)$ corresponds to the positive electrode. The lithium
behavior in the electrolyte is coupled with a diffusion model for the
evolution of the Li concentration $c_{\rm s}(x;r,t)$ in a generic
solid spherical electrode particle. Spherical symmetry of
this diffusion is assumed along the radial $r$--direction in the region $r\in
[0,R_{\rm s}]$ where $R_{\rm s}$ (m) is the average radius of a
generic particle. Using a 1D approximation for the entire battery is
valid since the characteristic length scale of a typical Li-ion cell
along the $x$-axis is on the order of 100 $\mu$m, whereas the
characteristic length scale for the remaining two axes is on the order
of 100,000 $\mu$m or more (cf. \cite{2010ChEtAl}). Note that $R_{\rm
  s}$ can be different in each electrode ($R_{\rm s,-}$, $R_{\rm
  s,+}$).

Based on the conservation equations deduced in Remark \ref{resce}, and
the models appearing in the literature
(cf. \cite{1993DoFuNe,1994FuDoNe,2002GoEtAl,2005FaPl,2006SmWa,2006-b-SmWa,2007SmRaWa,2010ChEtAl,2012KiSmIrPe},
along with the additional assumption of constant diffusion and
activity electrolyte coefficients, a model for the performance of a
battery at constant temperature, is given by system of equations
(\ref{eqce})--(\ref{eqps}):

\begin{equation} \label{eqce}
\left\{
\begin{array}{l}
{\displaystyle\varepsilon_{\rm e} \frac{\partial  c_{\rm e}}{\partial t} -
D_{\rm e} \frac{\partial }{\partial x} \left( \varepsilon_{\rm e}^p \frac{\partial c_{\rm e}}{\partial x} \right) = \frac{1-t^0_+}{F}
j^{\rm Li}, }
\hspace*{.2cm}\mbox{in } (0,L)\times (0,t_{\rm end}),
\\[.3cm]
{\displaystyle
\frac{\partial c_{\rm e}}{\partial x} (0,t) = \frac{\partial c_{\rm e}}{\partial x} (L,t) =0, \ \ \ t\in (0,t_{\rm end}),
} \\[.3cm]
{\displaystyle
c_{\rm e} (x,0) = c_{\rm e,0} (x), \ \ \ x\in (0,L),
}
\end{array}
\right.
\end{equation}

\begin{equation} \label{eqcs}
 \left\{
\begin{array}{l}
\mbox{For each }x\in (0,L_1) \cup (L_1+\delta,L): \\[.3cm]
{\displaystyle
\frac{\partial  c_{\rm s}}{\partial t} -
\frac{D_{\rm s}}{r^2} \frac{\partial }{\partial r} \left(  r^2 \frac{\partial c_{\rm s}}{\partial r} \right)
= 0, \ \mbox{ in } (0,R_{\rm s})\times (0,t_{\rm end}),
} \\[.3cm]
{\displaystyle \frac{\partial c_{\rm s}}{\partial r} (x;0,t) = 0, \ \  -D_{\rm s}\frac{\partial c_{\rm s}}{\partial r} (x;R_{\rm s},t) =
\frac{R_{\rm s}(x)}{3\varepsilon_{\rm s}(x)F}
j^{\rm Li} , \ \ t\in (0,t_{\rm end}),} \\[.3cm]
 c_{\rm s}(x;r,0)= c_{\rm s,0}(x;r),
\end{array}
\right.
\end{equation}

\begin{equation} \label{eqpe}
 \left\{
\begin{array}{l}
{\displaystyle \mbox{For each }t\in (0,t_{\rm end}):} \\[.3cm]
{\displaystyle
-\frac{\partial }{\partial x} \left( \varepsilon_{\rm e}^p\kappa \frac{\partial \phi_{\rm e}}{\partial x} \right)
+ (1-2t^0_+)\frac{R  T}{F}\frac{\partial }{\partial x} \left( \varepsilon_{\rm e}^p\kappa \frac{\partial}{\partial x}  \ln \big( c_{\rm e}\big) \right)
} = j^{\rm Li} \mbox{ in } (0,L),  \\[.3cm]
{\displaystyle \frac{\partial \phi_{\rm e}}{\partial x} (0,t) = \frac{\partial \phi_{\rm e}}{\partial x} (L,t) = 0},
\end{array}
\right.
\end{equation}

\begin{equation} \label{eqps}
 \left\{
\begin{array}{l}
{\displaystyle \mbox{For each }t\in (0,t_{\rm end}):} \\[.3cm]
{\displaystyle
-\varepsilon_{\rm s} \sigma \frac{\partial^2 \phi_{\rm s}}{\partial x^2}
= } -j^{\rm Li}  \ \mbox{ in } (0,L_1)\cup (L_1+\delta ,L),
 \\[.5cm]
{\displaystyle \varepsilon_{\rm s}(0)\sigma (0) \frac{\partial \phi_{\rm s}}{\partial x} (0,t) = \varepsilon_{\rm s}(L)\sigma (L) \frac{\partial \phi_{\rm s}}{\partial x}
(L,t) = -\frac{I(t)}{A},} \\[.3cm]
{\displaystyle \frac{\partial \phi_{\rm s}}{\partial x} (L_1,t) = \frac{\partial \phi_{\rm s}}{\partial x} (L_1+\delta ,t) = 0}, \\[.3cm]
\end{array}
\right.
\end{equation}
In this system of equations the independent variable are position $x$
(m) and time $t$ (s) with dependent variables $c_{\rm e}= c_{\rm
  e}(x,t)$ (mol m$^{-3}$), $c_{\rm s} = c_{\rm s} (x;r,t)$ (mol
m$^{-3}$), $\phi_{\rm e} = \phi_{\rm e} (x,t)$ (V), $\phi_{\rm s} =
\phi_{\rm s} (x,t)$ (V), and $I=I(t)$ (A) the applied current.
Parameters in the model are
$p$ the Bruggeman
porosity exponent (nondimensional constant), $D_{\rm e}$ the
electrolyte diffusion coefficient (m$^2$ s$^{-1}$), $t^0_+$ the transference
number of Li$^+$,
$\kappa= \kappa \left(c_{\rm
    e}(x,t)\right)$ the electrolyte phase ionic conductivity (S
m$^{-1}$), and  $A$ (m$^2$) is the cross-sectional area (also the current
collector area),
There
are parameters that take different values in different regions and
these are $\varepsilon_{\rm e}$ ($\varepsilon_{{\rm e},-}$ if $x\in
(0, L_1)$, $\varepsilon_{\rm e,sep}$ if $x\in (L_1, L_1+\delta)$ and
$\varepsilon_{{\rm e},+}$ if $x\in (L_1+\delta ,L)$) which is the
volume fraction of the electrolyte, $D_{\rm s}$ ($D_{\rm s,-}$ if $x\in (0, L_1)$,
$D_{\rm s,+}$ if $x\in (L_1+\delta ,L)$) which is the solid phase Li
diffusion coefficient (m$^2$ s$^{-1}$), $\varepsilon_{\rm s}$
($\varepsilon_{{\rm s},-}$ if $x\in (0, L_1)$, $\varepsilon_{{\rm
    s},+}$ if $x\in (L_1+\delta ,L)$) which is the volume fraction of the
active materials in the electrodes,
$\sigma_{\rm s}$ ($\sigma_{{\rm s},-}$ if $x\in (0,
L_1)$, $\sigma_{{\rm s},+}$ if $x\in (L_1+\delta ,L)$) which is the
electrical conductivity of solid active materials in an electrode (S
m$^{-1}$), and $ j^{\rm Li}$ (A m$^{-3}$)
is the reaction current resulting from intercallation of Li into solid
electrode particles.

For $j^{\rm Li}$ the following Butler-Volmer equation is commonly used
(cf. \cite{1993DoFuNe,2002GoEtAl,2006SmWa,2006-b-SmWa,2007SmRaWa,2010ChEtAl,2012KiSmIrPe})
\begin{equation} \label{fcbv}
%\hspace*{-.2cm}j^{\rm Li} \Big( x,\phi_{\rm s},\phi_{\rm e}, c_{\rm s}, c_{\rm e}\Big) =
\hspace*{-.2cm}j^{\rm Li} =
\left\{
\begin{array}{l}
 {\displaystyle \frac{3\varepsilon_{\rm s} (x) }{R_{\rm s}(x)} i_0  \left( \exp \left( \frac{\alpha_{\rm a}F}{R \ T}\eta \right) - \exp \left( \frac{-\alpha_{\rm c}F}{R \ T}\eta \right) \right) } \\[.4cm]
 \hspace*{2cm} \mbox{ if } x\in
 (0,L_1)\cup (L_1+\delta , L), \\[.4cm]
 0  \qquad \qquad \mbox{ if } x\in (L_1,L_1+\delta)
\end{array}
\right.
\end{equation}
(here, for the sake of simplicity, we have considered the solid/electrolyte interfacial film resistance to be zero and therefore is not included in the above equation), where
$T$ (K) is the temperature of the cell, $a_{\rm s}(x)=\frac{3\varepsilon_{\rm s} (x) }{R_{\rm s}(x)}$ (m$^{-1}$) is the specific interfacial area of electrodes, $i_0=i_0 (x,c_{\rm s},c_{\rm e})$ (A m$^{-2}$)
is the exchange current density of an electrode reaction
, $\alpha_{\rm a}$,
$\alpha_{\rm c}$ (dimensionless) are
anodic and cathodic transfer coefficients for an electrode reaction,
and $\eta$ (V) is the surface overpotential (V) of an electrode reaction.

Most of the parameters in the Butler-Volmer equation are agreed upon in models but there is considerable variation in the behavior of $i_0$ and $\eta$. Typically
%$$
%\eta=\eta \Big( x,\phi_{\rm s}(x,t),\phi_{\rm e}(x,t), c_{\rm s}(x;R_{\rm s}(x),t)\Big),
%$$
$$
\eta = \left\{
\begin{array}{l}
 {\displaystyle \phi_{\rm s} - \phi_{\rm e} - U(x,c_{\rm s}), \mbox{ if } \quad x\in
 (0,L_1)\cup (L_1+\delta , L),} \\[.2cm]
 0 \qquad \qquad \mbox{ if } x\in (L_1,L_1+\delta),
\end{array}
\right.
$$
where $U(x,c_{\rm s})$ is the equilibrium potential (V) at the
solid/electrolyte interface (i.e. the open circuit voltage, OCV, of
each electrode).  An empirical function is usually exploited for $U$
taking different forms in the two electrodes and dependent only on the
local surface concentration of lithium with
$$
 U =
\left\{
\begin{array}{ll}
 U_{-}\left( \displaystyle\frac{c}{c_{\rm s,-,max}}\right)
 & \mbox{ if } x\in (0, L_1), \\
 U_{+}\left( \displaystyle\frac{c}{c_{\rm s,+,max}}\right)
 & \mbox{ if } x\in (L_1+\delta ,L),
\end{array}
\right.\;.
$$
The functions $U_{-},U_{+}$ are typically obtained from fitting
experimental data and the constants $c_{\rm s,+,max}$, $c_{\rm s,-,max}$ (mol
m$^{-3}$) taken as the maximum possible concentration in the solid
positive and negative electrode, respectively.  The exchange current
$i_0$ is taken in different forms including being constant and being
\begin{equation} \label{ddio}
 i_0 =
\left\{
\begin{array}{ll}
  k_- \;c_{\rm e}^{\alpha_{\rm a}}\; (c_{\rm s,-,max} -c_{\rm s})^{\alpha_{\rm a}} \;c_{\rm s}^{\alpha_{\rm c}} & \mbox{ if } x\in (0, L_1), \\
  k_+ \;c_{\rm e}^{\alpha_{\rm a}} (c_{\rm s,+,max} -c_{\rm s})^{\alpha_{\rm a}} \;c_{\rm s}^{\alpha_{\rm c}} & \mbox{ if } x\in (L_1+\delta ,L),
\end{array}
\right.
\end{equation}
where $k_-,k_+$ are kinetic rate constants (A m$^{-2+6\alpha_{\rm a}+3\alpha_{\rm c}}$ mol$^{-2\alpha_{\rm a}-\alpha_{\rm c}}$).
In the final section we concentrate on the appropriateness of the functional form of $U$ and $i_0$.

\begin{remark} \label{semist} The second term on the left hand side of
  system (\ref{eqpe}) is often written in the literature, using
  $(1-t^0_+)$ (see \cite{1993DoFuNe}, \cite{1994FuDoNe}) or
  $2(1-t^0_+)$ (see \cite{2002GoEtAl}, \cite{2006SmWa},
  \cite{2006-b-SmWa}, \cite{2007SmRaWa}, \cite{2010ChEtAl},
  \cite{2012KiSmIrPe}), instead of $(1-2t^0_+)$, which is the
  term deduced in Remark \ref{resce}).
\end{remark}

\begin{remark} \label{rsfp} This system of equations does not have
  uniqueness of solution (if $\phi_{\rm s}(x,t)$ and $\phi_{\rm
    e}(x,t)$ are solutions then $\phi_{\rm s}(x,t)+c(t)$ and
  $\phi_{\rm e}(x,t)+c(t)$ are also solutions, for any function
  $c(t)$). A way of avoiding that is to set a reference value of
  $\phi_{\rm s}(x,t)$ or $\phi_{\rm e}(x,t)$ at some point $x$.  For
  instance we can impose $\phi_{\rm s} (0,t)=0$ for any $t\in
  [0,t_{\rm end}]$. Some results regarding the existence and
  uniqueness of solution can be seen in \cite{2014Ra}.
\end{remark}

\begin{remark}

After solving the above model, we can estimate the state of the charge of the negative electrode SOC$_-(t)$ and of the  positive one SOC$_+(t)$,
 and the cell voltage $V(t)$, at time $t$, by computing
$$
\mbox{SOC}_-(t)=\frac{3}{L_1(R_{\rm s-})^3} \int_0^{L_1}\int_0^{R_{\rm s,-}} r^2 \frac{c_{\rm s}(x;r,t)}{c_{\rm s,-,max}}{\rm d}r{\rm d}x,
$$
$$
\mbox{SOC}_+(t)=\frac{3}{(L-L_1-\delta)(R_{\rm s+})^3} \int_{L_1+\delta}^{L}
\int_0^{R_{\rm s,+}} r^2 \frac{c_{\rm s}(x;r,t)}{c_{\rm s,+,max}}{\rm d}r{\rm d}x,
$$
$$
 V(t) = \phi_{\rm s} (L,t) - \phi_{\rm s} (0,t) - \frac{R_{\rm f}}{A}I(t),
$$
where there is a constant film resistance of $R_{\rm f}$.
\end{remark}

\begin{remark} \label{rmswbc}
In \cite{2002GoEtAl},
\cite{2006SmWa},
\cite{2006-b-SmWa},
\cite{2007SmRaWa},
\cite{2012KiSmIrPe},
authors use the following (incorrect) boundary conditions at $x=0$ and $x=L$, instead of those presented in (\ref{eqps}):
\begin{equation} \label{wbc1}
 -\varepsilon_{\rm s,-}\sigma_- \frac{\partial \phi_{\rm s}}{\partial x} (0,t) = \varepsilon_{\rm s,+}\sigma_+ \frac{\partial \phi_{\rm s}}{\partial x} (L,t) = \frac{I(t)}{A}.
\end{equation}
We note in passing that if this incorrect condition is used then it can
be proved (see \cite{2014Ra}) that the corresponding system of boundary value problems
does not have any solution unless $I(t) \equiv 0$.
\end{remark}

\section{The Butler-Volmer equation} \label{sdbve}

The Butler-Volmer equation in the mathematical model presented here
has a general functional form widely used in the literature
(cf. \cite{1993DoFuNe,1994FuDoNe,2002GoEtAl,2006SmWa,2006-b-SmWa,2007SmRaWa,2010ChEtAl,2012KiSmIrPe}),
but it is necessary to take care in the detailed functions used for
$U$ and $\eta$ as explained below. We note that both $i_0$ and $U$ may
vary with $c_{\rm s}$ and $c_{\rm e}$ throughout the relevant range of these
concentrations but we are only interested in ensuring that the
behavior is appropriate when the electrodes get close to being fully
intercollated $c_{\rm s}=c_{\rm s,\pm,max}$, are fully depleted $c_{\rm s}=0 $, or
where the electrolyte is completely depleted $c_{\rm e}= 0$. To discuss
the possible behavior we consider the case $c_{\rm s} \to 0$ and then indicate
how this can be extended to the other cases.

Consider therefore the case $c_{\rm s} \to 0$ while $U$ tending to a
constant and $i_0$ is given by (\ref{ddio}) so that $i_0\to 0$. It might
be expected that the Li flux out of the particle will therefore cease
as $c_{\rm s}  \to 0$ thereby preventing negative concentrations in the
solid. However, this condition also implies that if we start with a
depleted particle then no flux can ever enter the particle. Such a
physically unrealistic situation should not be allowed in the model.

Alternative formulations of $i_0$ and $U$ can avoid this problem in
various ways and it is always possible to simply chose these functions
and then impose some switching logic to turn the flux on or off as
required to avoid such problems. However, it may be preferable to have
the Butler-Volmer condition designed, through the modeling of
suitable physical mechanisms, to ensure such physically irrelevant
situations cannot occur

Hence we would like to have the property that, as $c_{\rm s}  \to 0$ the
flux can only take negative values. Such a property can readily be
achieved by ensuring that $U\to \infty$ as $c_{\rm s}  \to 0$ as this will
ensure that the negative exponential dominates the Butler-Volmer
condition. A common physical condition is to consider that, as $c_{\rm s}  \to
0$ the forward (cathodic) should be of zeroth order in $c_{\rm s} $ while the reverse (anodic)
reaction should be first order. Such behavior can be readily achieved
by taking $i_0$ and $U$ to have the local form, when $c_{\rm s}  \to
0$,
$$
i_0\approx
\left(
  k\;c_{\rm s}\right)^{\left(\frac{\alpha_{\rm c}}{\alpha_{\rm a}+\alpha_{\rm c}}\right)}
\ \
\mbox{ and }
\ \
 U\approx
 \frac{-RT}{F(\alpha_{\rm a}+\alpha_{\rm c})}\ln \left( k\;c_{\rm s} \right),
$$
where $k$ is a positive constant (which can be different for each electrode and also can change inside non-homogeneous electrodes). Such a
formulation near this extreme of the concentration will automatically
ensure the flux cannot become positive at this limit of the
concentration reduces, thereby avoiding negative concentrations, and
the flux can be finite and positive in this limit so the particle can
be charged from a completely depleted state. Note that
a reverse reaction of order greater than one can be analyzed but
this is not usually considered.

Such local behavior of the Butler-Volmer condition is included in some
formulation but these need to include all possibilities including
extremely low electrolyte concentrations.

Therefore, a reasonable Butler-Volmer formulation would be
one where
\begin{equation} \label{fdi0}
i_0 (x,c_{\rm s},c_{\rm e})=
i_{\rm a}(x,c_{\rm s})^{\frac{\alpha_{\rm c}}{\alpha_{\rm a}+\alpha_{\rm c}}}
i_{\rm c}(x,c_{\rm e},c_{\rm s})^{\frac{\alpha_{\rm a}}{\alpha_{\rm a}+\alpha_{\rm c}}}
\end{equation}
and
\begin{equation} \label{fdU}
 U(x,c_{\rm s},c_{\rm e})=
 \frac{RT}{F(\alpha_{\rm a}+\alpha_{\rm c})}\ln \left( \frac{i_{\rm c}(x,c_{\rm e},c_{\rm s})}{ i_{\rm a}(x,c_{\rm s})} \right).
\end{equation}
The functions $i_{\rm a}$ and $i_{\rm c}$ must be strictly positive except at the extremes of the concentrations and locally these functions must take the form:\\
\hspace*{0.5cm} i) as $c_{\rm s} \to 0 $
$$
i_{\rm a}(x,c_{\rm s}) \approx
  k_{\rm a}(x) \;c_{\rm s} ,
$$
with $k_{\rm a}(x)= k_{\rm a,+}$ or $k_{\rm a,-}$, if $x\in (0,L_1)$ or $x\in( L_1+\delta ,L)$, respectively (again, non-homogeneous electrodes could also be considered).
\\
\hspace*{0.5cm}
ii)
as $c_{\rm s} \to c_{\rm s,\pm,max}$
$$
i_{\rm c} (x,c_{\rm e},c_{\rm s}) \approx
  k_{\rm c,s}(x,c_{\rm e})\; (c_{\rm s,\pm,max}-c_{\rm s}) .
$$
and\\
\hspace*{0.5cm}
iii) as $c_{\rm e} \to 0 $
$$
i_{\rm c} (x,c_{\rm e},c_{\rm s}) \approx
  k_{\rm c,e}(x,c_{\rm s})\; c_{\rm e} .
$$
Note there may also be a need to consider a maximum electrolyte
concentration to avoid precipitation and this might be accommodated in
a similar manner.

Notice four important features of the approach proposed here in
contrast with what is commonly found in the literature:
\begin{enumerate}
\item The function $U$, the OCV, also depends on $c_{\rm e}$ (instead
  of being independent of it).
 \item If $c_{\rm s}\rightarrow 0$, then $i_0\rightarrow 0$ and $U\rightarrow \infty$ (instead of tending to a finite value) so that
 $$
j^{\rm Li} \approx \frac{3\varepsilon_{\rm s}}{R_{\rm s}}
\left(
B_1 c_{\rm s} \exp \left( \frac{\alpha_{\rm a}F}{R \ T}(\phi_{\rm s}-\phi_{\rm e}) \right)-D_1\exp \left( \frac{-\alpha_{\rm c}F}{R \ T}(\phi_{\rm s}-\phi_{\rm e}) \right)
\right)
  $$
for some strictly positive constants $B_1$ an $D_1$.
  \item If $c_{\rm e}\rightarrow 0$, then $i_0\rightarrow 0$ and $U\rightarrow -\infty$ so that
 $$
j^{\rm Li} \approx \frac{3\varepsilon_{\rm s}}{R_{\rm s}}
\left(
B_2 \exp \left( \frac{\alpha_{\rm a}F}{R \ T}(\phi_{\rm s}-\phi_{\rm e}) \right)
-D_2 c_{\rm e}\exp \left( \frac{-\alpha_{\rm c}F}{R \ T}(\phi_{\rm s}-\phi_{\rm e}) \right)
\right)
  $$
for some strictly positive constants $B_2$ an $D_2$.
\item
If $c_{\rm s}\rightarrow c_{\rm s,\pm,max}$, then
  $i_0\rightarrow 0$ and $U\rightarrow -\infty$
and
$$
j^{\rm Li} \approx \frac{3\varepsilon_{\rm s}}{R_{\rm s}}
\left(
B_3 \exp \left( \frac{\alpha_{\rm a}F}{R \ T}(\phi_{\rm s}-\phi_{\rm e}) \right) \right. \hspace*{4cm}
$$
$$
\hspace*{4cm} \left. -D_3 (c_{\rm s}-c_{\rm s,\pm,max})\exp \left( \frac{-\alpha_{\rm c}F}{R \ T}(\phi_{\rm s}-\phi_{\rm e}) \right)
\right)
  $$
for some strictly positive constants $B_3$ an $D_3$.
\end{enumerate}
These conditions ensure that the lithium flux is constrained to
prevent concentrations in the solid being taken into an unphysical
regime and also ensuring that, if the solid is near such extreme
conditions, the flux can move the system away from these. This ensures
the battery can readily charge from a completely depleted state or discharge
from a fully charged state.

\begin{remark}
  The most simple case of functions $i_0$ and $U$ satisfying the above conditions (of course, many other possibilities can be also considered) is obtained with
  $$
i_{\rm a}(x,c_{\rm s}) =
\left\{
\begin{array}{ll}
  k_{\rm a,-} c_{\rm s} & \mbox{ if } x\in (0, L_1), \\
  k_{\rm a,+} c_{\rm s} & \mbox{ if } x\in (L_1+\delta ,L),
\end{array}
\right.
$$

$$
i_{\rm c}(x,c_{\rm s},c_{\rm e}) =
\left\{
\begin{array}{ll}
  k_{\rm c,-} c_{\rm e}(c_{\rm s,-,max}-c_{\rm s}) & \mbox{ if } x\in (0, L_1), \\
  k_{\rm c,+} c_{\rm e}(c_{\rm s,+,max}-c_{\rm s}) & \mbox{ if } x\in (L_1+\delta ,L).
\end{array}
\right.
$$
Therefore, according to (\ref{fdU}),
$$
 U(x,c_{\rm s},c_{\rm e})
=
\left\{
\begin{array}{ll}
 \frac{RT}{F(\alpha_{\rm a}+\alpha_{\rm c})}\ln \left( \frac{k_{\rm c,-}c_{\rm e}(c_{\rm s,-,max}-c_{\rm s})}{ k_{\rm a,-}c_{\rm s}} \right)
 & \mbox{ if } x\in (0, L_1), \\
 \frac{RT}{F(\alpha_{\rm a}+\alpha_{\rm c})}\ln \left( \frac{k_{\rm c,+}c_{\rm e}(c_{\rm s,+,max}-c_{\rm s})}{ k_{\rm a,+}c_{\rm s}} \right)
 & \mbox{ if } x\in (L_1+\delta ,L).
\end{array}
\right.
$$
Furthermore, according to (\ref{fdi0}),
$$
i_0 (x,c_{\rm s},c_{\rm e}) =
\left\{
\begin{array}{ll}
  k_{\rm -} c_{\rm e}^{\frac{\alpha_{\rm a}}{\alpha_{\rm a}+\alpha_{\rm c}}} \; (c_{\rm s,-,max}-c_{\rm s})^{\frac{\alpha_{\rm a}}{\alpha_{\rm a}+\alpha_{\rm c}}} \; c_{\rm s}^{\frac{\alpha_{\rm c}}{\alpha_{\rm a}+\alpha_{\rm c}}} & \mbox{ if } x\in (0, L_1), \\
  k_{\rm +} c_{\rm e}^{\frac{\alpha_{\rm a}}{\alpha_{\rm a}+\alpha_{\rm c}}} \; (c_{\rm s,+,max}-c_{\rm s})^{\frac{\alpha_{\rm a}}{\alpha_{\rm a}+\alpha_{\rm c}}}
  \; c_{\rm s}^{\frac{\alpha_{\rm c}}{\alpha_{\rm a}+\alpha_{\rm c}}}
  & \mbox{ if } x\in (L_1+\delta ,L),
\end{array}
\right.
$$
with $k_-=k_{\rm a,-}^{\alpha_{\rm c}} k_{\rm c,-}^{\alpha_{\rm a}}$ and $k_+=k_{\rm a,+}^{\alpha_{\rm c}} k_{\rm c,+}^{\alpha_{\rm a}}$, which coincides with (\ref{ddio}) when $\alpha_{\rm a}+\alpha_{\rm c} =1$.
\end{remark}

\begin{remark}
  Some authors
  (cf. \cite{2006SmWa,2006-b-SmWa,2007SmRaWa,2012KiSmIrPe})) take a
  constant value for $i_0$ and some of them (cf. \cite{2006SmWa})
  claim that it exhibits modest dependency on electrolyte and solid
  surface concentration. Although this can be valid for appropriate
  particular cases, in a general situation this does not seem to be
  valid, since $i_0$ may vary importantly, with extreme cases when the
  battery is either fully charged or fully discharged.
\end{remark}

\section{Conclusions}

The Butler-Volmer equation is commonly used in the literature in order
to take into account the electrochemical reactions that take place in
battery electrodes. It is widely used in mathematical models for the
simulation of a Lithium-ion battery, based on a macro-homogeneous
approach developed by Neuman \cite{1973Ne}. It has been shown that the
way this equation is commonly used in these models may create
situations where the system becomes non physical. Conditions on the
functional form of the Butler-Volmer equation that ensure this will not
occur have been presented.  Furthermore, some common mistakes that can
be found in the literature regarding the mathematical equations of the
Neuman-type models have also been pointed out.

\section*{Acknowledgments}

The authors wish to thank Prof. S.J. Chapman for his many helpful
comments and suggestions.  This work was carried out thanks to the
financial support of the Spanish Ministry of Education, Culture and
Sport; the Ministry of Economy and Competitiveness under project
MTM2011-22658; the "Junta de Andaluc\'{\i}a" through project
P12-TIC301; and the research group MOMAT (Ref. 910480) supported by
''Banco de Santander" and ''Universidad Complutense de Madrid". This
publication was also supported by OCIAM (University of Oxford), where
the first author was collaborating as an academic visitor.

\bibliographystyle{elsarticle-num}
\bibliography{bib-database}






% \end{thebibliography}

\end{document}